\documentclass[a4paper,11pt]{article}
\usepackage[utf8]{inputenc}
\usepackage[dvipsnames]{xcolor}
\usepackage{amsmath,mathtools,mathrsfs,amsfonts,physics,bm,empheq,tikz,graphicx,epstopdf,tabu,color,enumerate,pifont}
\usepackage{pgfplots,float,textgreek}

\pgfplotsset{
compat=1.10,
every axis plot/.append style={no marks,thick},
every axis/.style={
  axis lines=middle,
  width=10cm,
  height=7cm,
  }
}

\usepackage[percent]{overpic}

\renewcommand{\Psi}{\text{\textpsi}}
\renewcommand{\nu}{\text{\textnu}}
\renewcommand{\pi}{\text{\textpi}}
\renewcommand{\Pi}{\text{\textPi}}

\usepackage[left=20mm, top=25mm, right=20mm, bottom=25mm]{geometry}

\usepackage{environ}

\newcommand{\bb}[1]{\pmb{#1}}

\NewEnviron{myequation}{%
\begin{equation}
\scalebox{1.2}{$\BODY$}
\end{equation}
}

\makeatletter
\newcommand*{\defeq}{\mathrel{\rlap{%
                     \raisebox{0.3ex}{$\m@th\cdot$}}%
                     \raisebox{-0.3ex}{$\m@th\cdot$}}%
                     =}
\makeatother

\title{\textbf{Quantum First Passage Time Problem}\\\large--A Bohmian Perspective\footnote{This paper is based on my talk presented at the winter seminar of the research group ``Mathematical Foundations of Physics" of LMU Munich, which took place at Wunsiedel, a small town in northeastern Bavaria during the last week of February 2017.}}
\author{Siddhant Das\footnote{\texttt{Siddhant.Das@physik.uni-muenchen.de}}}
\date{%
\small Elite master program `Theoretical and Mathematical Physics'\\Arnold Sommerfeld Center, Ludwig-Maximilians-Universit\"{a}t M\"{u}nchen, Germany.\\[2ex]%
\today}

\begin{document}
\maketitle

\begin{abstract}
\noindent The prediction of arrival time or first passage time statistics of a quantum particle is an open problem, which challenges the foundations of quantum theory. One of the most promising and insightful approaches to this problem stems from the de Broglie-Bohm pilot-wave theory (a.k.a Bohmian mechanics). Applying the fundamental postulates of this theory, we analyze a simplified first passage time experiment and derive the empirical passage time distribution $\Pi(\tau)$. Implications of our results are also discussed.\\[7pt]Keywords: arrival time operator, tunneling time, Bohmian mechanics, pilot-wave theory
\end{abstract}
\section{Introduction}
In non-relativistic quantum mechanics, the probability of finding a particle in a small spatial volume $\mathrm{d}^3r$ around position $\bb{r}$ at a fixed time $t$ is given by Born's rule $|\psi(\bb{r},t)|^2\mathrm{d}^3r$, where $\psi(\bb{r},t)$ is the wave function of the particle. However, a formula for the probability of finding the particle at a fixed point $\bb{r}$ between times $t$ and $t+\mathrm{d}t$ is the matter of an ongoing debate. One might wonder, why is it so easy to speak about a position measurement at a fixed time, yet so hard to speak about a time measurement at a fixed position? For concreteness, consider the following experiment: a particle of mass $m$ is prepared in the state\footnote{State preparation is discussed in detail below.}
\begin{equation}\label{init}
    \psi_0(\bb{r})=\frac{e^{-\frac{1}{2}(r/a)^2}}{(\sqrt{\pi}\,a)^{3/2}}
\end{equation}
at time $t=0$, where $a$ is some fixed width of the wave function and $\bb{r}\equiv(r,\vartheta,\phi)$ are the standard spherical polar coordinates. A spherical detector placed at $r=d$ registers a click when the particle crosses $r=d$ and records the \emph{first passage time} of the particle, denoted by $\tau$. Let's assume that the experiment can be repeated several times, keeping $\psi_0$ unchanged in each run. Unsurprisingly, the detector click instants would vary from experiment to experiment, i.e., one would obtain a random sequence of passage times $\tau_1$, $\tau_2$, $\tau_3\,...$ What is the probability distribution of these passage times $\Pi(\tau)$ as a function of $a$ and $d$?

The prediction of first passage times of a quantum particle has a long history \cite{MUGA,MUGA1}. The very notion of arrival or passage time of a particle is not well posed within the orthodox (or Copenhagen) interpretation of quantum mechanics, since the particle is said to not have a well defined position at a given instant of time. However, the problem of timing the motion of quantum particles surfaced long before any known interpretations of quantum mechanics came into being. As early as 1925, shortly after the invention of matrix mechanics, Wolfgang Pauli wrote to Niels Bohr:\\[10pt]
\noindent``\textit{In the new theory, all physically observable quantities still do not really occur. Absent, namely, are the time instants of transition processes, which are certainly in principle observable (as for example, are the instants of the emission of photoelectrons). It is now my firm conviction that a really satisfying physical theory must not only involve no unobservable quantities, but must also connect all observable quantities with each other. Also, I remain convinced that the concept of `probability' should not occur in the fundamental laws of a satisfying physical theory.}''(\S\,1.1 \cite{MUGA})\\

\noindent Pauli's early views on the problem of time in quantum mechanics greatly influenced the subsequent research on this subject. In particular, he showed that a self-adjoint time operator $\hat{T}$, canonically conjugate to the Hamiltonian $\hat{H}$, viz.
\begin{equation}
    [\hat{H},\hat{T}]=i\hbar
\end{equation}
(just like position and momentum) implied that the spectrum of $\hat{H}$ would be \emph{unbounded} from below, which in turn implied that matter couldn't be stable. This result raised doubts on the status of the `time-energy uncertainty relation'
\begin{equation}
    \Delta E\,\Delta T\geq\frac{\hbar}{2}.
\end{equation}
Despite these impediments, many physicists have attempted to incorporate a respectable time observable by extending the basic framework of quantum theory (see \cite{MUGA,MUGA1} for various attempts), although there is no general consensus among physicists on this subject.
\par The notion of arrival or first passage time is most naturally connected with that of particle trajectories, an idea not taken seriously, for example, in the Copenhagen interpretation of quantum mechanics. Therefore, it has long been realized that quantum theories comprising of actual particle trajectories, such as Bohmian Mechanics (a.k.a. de Broglie-Bohm pilot-wave theory, or the causal interpretation of quantum mechanics) provide a natural framework for addressing this problem.
\par In this theory, the idea of a particle is taken seriously, i.e., it is described as a point mass with a well defined trajectory $\bb{R}(t)$. The motion of the particle is choreographed by the wave function $\psi$, which satisfies the time dependent Schr\"{o}dinger equation \cite{Durr,Holland,BohmHiley}. Guided by the wave function, the particle executes a highly non-Newtonian motion (hence the name pilot-wave), which underlies the wave-like properties seen in interference experiments. The theory is deterministic, hence the characteristic randomness of quantum mechanical experiments is understood as an artifact of one's ignorance of initial conditions. Bohmian mechanics is shown to be empirically equivalent to quantum mechanics in the sense that it makes the same predictions as orthodox quantum mechanics, whenever the latter is unambiguous \cite{Durr,Holland}.
\par However, as indicated above, time measurements are problematic within the current formulation of quantum mechanics (also evidenced by recent attoclock experiments \cite{Keller,Lundsmann}). While there are a number of conflicting definitions of transit times, arrival times, etc., within orthodox quantum mechanics \cite{MUGA,MUGA1}, Bohmian mechanics privileges one, namely, the time taken by the Bohmian trajectory of the particle to strike a detector. Therefore, we ask: can the de-Broglie Bohm particle law of motion be made relevant to experiments?
We try to answer this question in this paper, using the above experiment as a prototype. However, ``we do not contest the correctness of quantum mechanics in the domain where it is \emph{unambiguous}, testable and confirmed, but enquire whether that domain can be enlarged''(\S\,5.5 \cite{Holland}).

\section{Elements of Bohmian mechanics}\label{theory}
In Bohmian mechanics a particle has a well defined position $\bb{R}(t)$ at time $t$, which is a vector in $\mathbb{R}^3$. In the course of time, the particle moves on a deterministic (Bohmian) trajectory $\bb{R}$ with velocity vector $\dot{\bb{R}}$ specified by the guidance law
\begin{equation}\label{guide}
    \dot{\bb{R}}(t)=\frac{\mathrm{d}}{\mathrm{d}t}\bb{R}(t)=\frac{\hbar}{m}\,\mathrm{Im}\!\left[\frac{\bb{\nabla}\psi}{\psi}\right]\!(\bb{R}(t),t).
\end{equation}
Here, $\psi(\bb{r},t)\in\mathbb{C}$ is the wave function of the particle, which satisfies the Schr\"{o}dinger equation
\begin{equation}\label{wave}
    i\hbar\frac{\partial}{\partial t}\psi(\bb{r},t)=-\frac{\hbar^2}{2m}\nabla^2\psi(\bb{r},t)+V(\bb{r})\psi(\bb{r},t)
\end{equation}
with some initial condition $\psi(\bb{r},0)\equiv\psi_0(\bb{r})$. Equations \eqref{guide} and \eqref{wave} describe an isolated spin-0 particle of mass $m$ and have analogues suitable for describing both multi-particle systems and particles with spin. The equations of motion in the latter cases are rather involved, and for simplicity will not be discussed here. The dynamical equations stated here are time reversal invariant, rotationally invariant, and the r.h.s. of \eqref{guide} transforms as a velocity under Galilean boosts. These properties qualify Bohmian mechanics as a legitimate nonrelativistic theory. Applying Eq. \eqref{guide} to any phenomena of interest, one obtains a very intuitive understanding of the actual dynamical processes at work, which are otherwise denigrated in the operational `shut up and calculate' approaches (see \cite{Norsen,Ward,Ward1,HDK,Colijn,Colijn1,Colijn2,Timko} for many detailed examples). A satisfactory account of the theory can be found in \cite{Durr,Holland,BohmHiley}.
\par As stated before, the probabilistic character of quantum mechanics arises in this theory as a consequence of ignorance of initial conditions, therefore the key insight for analyzing Bohmian mechanics lies within the foundations of statistical mechanics (especially in the ideas of Ludwig Boltzmann). This gives rise to the well known Born's rule \cite{DGZ}, which states that the particle position at time $t$ \emph{is} distributed according to $|\psi(\bb{r},t)|^2$, independent of any measurement prescription. In the next section we apply these basic principles to derive the empirical first passage time distribution $\Pi(\tau)$ for the experiment outlined above.

\section{Formulation}\label{example}
The picture we have in mind is that the wave function $\psi(\bb{r},t)$ evolves in time satisfying Eq. \eqref{wave} with initial condition $\psi_0$, while the particle moves on a well defined Bohmian trajectory $\bb{R}(t)$ satisfying \eqref{guide}, hence it's first passage time is unambiguously determined. One also needs to specify the initial position of the particle on the trajectory, viz. $\bb{R}(0)\equiv\bb{R}_0$ for solving Eq. \eqref{guide}. However, the particular $\bb{R}_0$ realized in an experiment is not known, hence the exact trajectory of the particle changes from experiment to experiment, and as a result the measured passage times appear random.
\par The first passage time of the particle is simply the first instant at which it's trajectory crosses $r=d$. More formally, we can write
\begin{equation}\label{defn}
    \tau(\bb{R}_0)=\mathrm{min}\{t|\,R(t,\bb{R}_0)=d,~\bb{R}_0\!\in\!\mathbb{R}^3\},
\end{equation}
where $R(t,\bb{R}_0)=\Vert\bb{R}(t,\bb{R}_0)\Vert$ is the radial coordinate of the particle at time $t$. We have explicitly indicated that the first passage time on any trajectory depends on the initial position $\bb{R}_0$. However, definition \eqref{defn} is incomplete, as it is not applicable to trajectories that \emph{never} cross $r=d$. For such trajectories, we can set the passage time to $\infty$, since in these instances the detector would never click.
\par We return to a more detailed description of the experiment and the results. Introducing new dimensionless variables
\begin{equation}\label{dim}
    \bb{r}'=\frac{\bb{r}}{a},\quad d'=\frac{d}{a},\quad\bb{R}'=\frac{\bb{R}}{a},\quad\psi'=\frac{\psi}{a^{-3/2}},\quad t'=\frac{\hbar}{ma^2}t ,
\end{equation}
we can rewrite the dynamical equations in a convenient nondimensionalized form, viz.,
\begin{align}
    \frac{\mathrm{d}}{\mathrm{d}t'}\bb{R}'(t')=\mathrm{Im}\!\left[\frac{\bb{\nabla}'\psi'}{\psi'}\right]\!(\bb{R}'(t'),t'),\label{guide1}\\ i\frac{\partial}{\partial t'}\psi'(\bb{r}',t')=-\frac{1}{2}\nabla'^{\,2}\psi'(\bb{r}',t'),\label{wave1}
\end{align}
where $\bb{\nabla}'$ denotes the gradient w.r.t. the primed coordinates, and the external potential $V$ has been set to zero. Henceforth, we will suppress the primes for brevity. 
\par We begin by solving the time dependent Schr\"{o}dinger equation \eqref{wave1} with initial condition $\psi_0(\bb{r})=\pi^{-3/4}e^{-r^2/2}$. An easy way to accomplish this is by means of Fourier transforms. Employing standard Fourier transform conventions
\begin{subequations}
\begin{align}
    \tilde{\psi}(\bb{k},t)&=\int_{\mathbb{R}^3}\!\!\mathrm{d}^3r\,e^{-i\,\bb{k}\cdot\bb{r}}\psi(\bb{r},t),\\\psi(\bb{r},t)&=\int_{\mathbb{R}^3}\!\frac{\mathrm{d}^3k}{(2\pi)^3}\,e^{i\,\bb{k}\cdot\bb{r}}\tilde{\psi}(\bb{k},t),\label{back}
\end{align}
\end{subequations}
we substitute Eq. \eqref{back} into \eqref{wave1}, obtaining
\begin{equation}\label{psitildek}
    \frac{\partial}{\partial t}\tilde{\psi}(\bb{k},t)=-i\frac{k^2}{2}\tilde{\psi}(\bb{k},t)\Rightarrow\tilde{\psi}(\bb{k},t)=A(\bb{k})e^{-i\frac{k^2}{2}t}.
\end{equation}
Here, $A(\bb{k})$ is an arbitrary function of $\bb{k}$, which can be determined from the initial condition $\psi_0$. In particular,
\begin{align}\label{four}
    A(\bb{k})=\tilde{\psi}(\bb{k},0)=\int_{\mathbb{R}^3}\!\!\mathrm{d}^3r\,e^{-i\,\bb{k}\cdot\bb{r}}\psi_0(\bb{r})=\pi^{-3/4}\int_{\mathbb{R}^3}\!\!\!\mathrm{d}^3r\,e^{-i\,\bb{k}\cdot\bb{r}-r^2/2}=(2\sqrt{\pi})^{3/2}e^{-k^2/2},
\end{align}
where the integral is easily evaluated in Cartesian coordinates. Substituting \eqref{four} into \eqref{psitildek}, and the result into \eqref{back}, we obtain the time dependent wave function
\begin{align}\label{soln}
    \psi(\bb{r},t)=(2\sqrt{\pi})^{3/2}\int_{\mathbb{R}^3}\!\frac{\mathrm{d}^3k}{(2\pi)^3}\,e^{i\,\bb{k}\cdot\bb{r}-(1+it)\frac{k^2}{2}}=\frac{e^{-\frac{r^2}{2(1+it)}}}{(\sqrt{\pi}\,(1+it))^{3/2}},
\end{align}
where the evaluation of the integral proceeds exactly as in \eqref{four}. From \eqref{soln} we see that the wave function propagates dispersively, i.e., it spreads isotropically in all directions with a width $\sigma(t)=\sqrt{1+t^2}$ that increases with time.\footnote{By width we mean that of $|\psi|=\sqrt{\psi\,\psi^*}$.}
\par Next, we look at the Bohmian trajectories, which are the integral curves of the Bohmian velocity field
\begin{equation}\label{v}
    \bb{v}_{\texttt{Bohm}}(\bb{r},t)=\mathrm{Im}\!\left[\frac{\bb{\nabla}\psi}{\psi}\right]\!(\bb{r},t)=\frac{t}{1+t^2}r\,\hat{\bb{r}},
\end{equation}
which in our case turns out to be a radial vector field. The particle position at time $t$ is
\begin{equation}
    \bb{R}(t)=R(t)\left[\cos\Theta(t)\sin\Phi(t)\,\hat{\bb{x}}+\sin\Theta(t)\sin\Phi(t)\,\hat{\bb{y}}+\cos\Phi(t)\,\hat{\bb{z}}\right],
\end{equation}
the time derivative of which is
\begin{equation}
    \dot{\bb{R}}(t)=\dot{R}(t)\,\hat{\bb{r}}(t)+R(t)\dot{\Theta}(t)\sin\Phi(t)\,\skew{4}{\hat}{\bb{\vartheta}}(t)+R(t)\dot{\Phi}(t)\,\skew{4}{\hat}{\bb{\phi}}(t),
\end{equation}
where
\begin{subequations}
\begin{align}
    \hat{\bb{r}}(t)&=\cos\Theta(t)\sin\Phi(t)\,\hat{\bb{x}}+\sin\Theta(t)\sin\Phi(t)\,\hat{\bb{y}}+\cos\Phi(t)\,\hat{\bb{z}},\\
    \skew{4}{\hat}{\bb{\vartheta}}(t)&=-\sin\Theta(t)\,\hat{\bb{x}}+\cos\Theta(t)\,\hat{\bb{y}},\\
    \skew{4}{\hat}{\bb{\phi}}(t)&=\cos\Theta(t)\cos\Phi(t)\,\hat{\bb{x}}+\sin\Theta(t)\cos\Phi(t)\,\hat{\bb{y}}-\sin\Phi(t)\,\hat{\bb{z}}.
\end{align}
\end{subequations}
The r.h.s of the guidance law \eqref{guide1} can be evaluated using Eq. \eqref{v}, and comparison with the above derivative yields the component equations
\begin{equation}
    \dot{R}(t)=\frac{t}{1+t^2}R(t),\quad R(t)\sin\Phi(t)\dot{\Theta}(t)=0,\quad R(t)\dot{\Phi}(t)=0.
\end{equation}
If we let $R(t)=0$, then all equations are trivially satisfied, however the initial condition $R(0)=R_0$ cannot be satisfied. For the same reason we cannot have $\sin\Phi(t)=0$, hence the only remaining possibility is $\dot{\Phi}(t)=0$ and $\dot{\Theta}(t)=0$. These equations are readily solved:
\begin{equation}
    \Theta(t)=\Theta_0,\quad \Phi(t)=\Phi_0,
\end{equation}
which imply that the particle moves radially on a straight line. The differential equation for the radial coordinate is separable and admits a simple solution of the form
\begin{equation}\label{rad}
    R(t)=R_0\sqrt{1+t^2}.
\end{equation}
Since $R(t)>0$ for all $t$, the particle moves radially outwards with a \emph{nonuniform} radial velocity (compare this with free Newtonian motion). However, as $t\to\infty$
\begin{equation}
    R(t) = R_0 t\,\sqrt{1 + \dfrac{1}{t^2}} = R_0 t \left(1 + \frac{1}{2t^2} - \dotsi \right) \sim R_0t+\mathcal{O}(t^{-1}),
\end{equation}
thus the velocity approaches $R_0$. This asymptotic radial velocity (restoring $\hbar$, $m$ and $a$),
\begin{equation}
    v_{\infty}=\frac{\hbar R_0}{ma^2},
\end{equation}
is a characteristic feature of free Bohmian motion. Note that particles starting far away from the centre of the wave packet acquire larger asymptotic velocities. In fact, all particles starting outside a sphere of radius $a^2/\text{\textlambda}_{\texttt{c}}$ acquire superluminal speeds as $t\to\infty$, where $\text{\textlambda}_{\texttt{c}}=\hbar/mc$ is the (reduced) Compton wavelength. This shouldn't come as a surprise, since Eq. \eqref{guide1} and \eqref{wave1} (just like Newton's equations of motion) are only Galileian covariant, hence do not comply with the principles of special relativity.
\par Since the Bohmian trajectories propagate radially outward, any trajectory crosses $r=d$ only once, provided $R_0<d$. The first passage time (or simply the passage time) $\tau$ can thus be determined by solving the equation $R(\tau)=d$, which yields (cf Eq. \eqref{defn})
\begin{equation}\label{time}
    \tau(\bb{R}_0)=\begin{cases}
       \sqrt{(d/R_0)^2-1}~&R_0\leq d\\
       \infty~&R_0>d
\end{cases}.
\end{equation}
We have set $\tau=\infty$ for all trajectories starting outside the detector, as they would not cross the detector in finite time. It must be remarked that Eq. \eqref{time} is exclusive to Bohmian mechanics with no known analogue in standard quantum mechanics. However, experience shows that in most situations of interest, the guidance law cannot be integrated analytically, hence an explicit formula connecting the passage time $\tau$ to the initial particle position $\bb{R}_0$, such as \eqref{time}, cannot be found. Thus, one can typically at best approximate $\Pi(\tau)$ from a large number of Bohmian trajectories, which must be {\emph {computed numerically}}.
\par We focus now on a derivation of the empirical passage time distribution $\Pi(\tau)$ for the case at hand. Recall that we are considering an ensemble of identically prepared experiments with $|\psi_0|^2$-distributed initial particle positions, hence the probability distribution of the passage time $\tau(\bb{R}_0)$ is given by
\begin{equation}\label{singular}
    \Pi(\tau)=\int_{\mathbb{R}^3}\!\!\!\mathrm{d}^3R_0~|\psi_0(\bb{R}_0)|^2\,\delta(\tau(\bb{R}_0)-\tau),
\end{equation}
where $\delta(x)$ is the Dirac delta function. Since $\tau(R_0)=\infty$ whenever $R_0>d$, one has to cautiously deal with the object $\delta(\infty-\tau)$. This calls for a long mathematical digression, which is not absolutely necessary, for we may instead consider the statistics of a quantity related to $\tau$, namely, its reciprocal
\begin{equation}\label{defnu}
    \nu(\bb{R}_0)=\tau^{-1}(\bb{R}_0)=\begin{cases}
       \frac{1}{\sqrt{(d/R_0)^2-1}}~&R_0\leq d\\
       0~&R_0>d
\end{cases}.
\end{equation}
Note that the distribution of the first passage time $\tau$, and that of it's reciprocal $\nu$, are equivalent statistical characterizations, hence there is no loss of generality in analyzing the latter. We do so, of course, because it is more amenable to mathematical analysis. Analogous to Eq. \eqref{singular}, the distribution of the reciprocal passage time $\nu(\bb{R}_0)$ can be written as
\begin{align}\label{lamb}
    \Lambda(\nu)=\int_{\mathbb{R}^3}\!\!\!\mathrm{d}^3R_0~|\psi_0(\bb{R}_0)|^2\,\delta(\nu(\bb{R}_0)-\nu)=4\pi\int_0^{\infty}\!\!\!\mathrm{d}R_0~R_0^2\,|\psi_0(\bb{R}_0)|^2\,\delta(\nu(\bb{R}_0)-\nu),
\end{align}
where the factor of $4\pi$ results from integrating over the angular coordinates of $\bb{R}_0$. Substituting $|\psi_0(\bb{R}_0)|^2=\pi^{-3/2}e^{-R_0^2}$, and Eq. \eqref{defnu}, we arrive at
\begin{equation}\label{split}
    \Lambda(\nu)=\frac{4}{\sqrt{\pi}}\int_0^d\!\!\mathrm{d}R_0~R_0^2\,e^{-R_0^2}\,\delta\!\left(\frac{1}{\sqrt{(d/R_0)^2-1}}-\nu\right)+\frac{4}{\sqrt{\pi}}\,\delta(\nu)\!\int_d^{\infty}\!\!\!\mathrm{d}R_0~R_0^2\,e^{-R_0^2}.
\end{equation}
The second integral multiplying $\delta(\nu)$, which we denote by $\alpha(d)$, can be evaluated relatively easily:
\begin{align}\label{alpha}
    \alpha(d) = \frac{4}{\sqrt{\pi}}\int_d^{\infty}\!\!\!\mathrm{d}R_0~R_0^2\,e^{-R_0^2}&=-\frac{2}{\sqrt{\pi}}\int_d^{\infty}\!\!\!\mathrm{d}\Big(e^{-R_0^2}\Big)R_0\nonumber\\&=-\frac{2}{\sqrt{\pi}}\left.R_0\,e^{-R_0^2}\,\right|_d^{\infty}+\frac{2}{\sqrt{\pi}}\int_d^{\infty}\!\!\!\mathrm{d}R_0\,e^{-R_0^2}\qquad(\text{integrating by parts})\nonumber\\&=\frac{2d}{\sqrt{\pi}}e^{-d^2}+\mathrm{erfc}(d),
\end{align}
where $\mathrm{erfc}(x)$ is the complementary error function (see Eq. (2.1.6) of \cite{Lebedev}). In retrospect, we see that $\alpha$ is simply the probability of finding the particle outside the detector at $t=0$.\footnote{In real experiments the initial wave function is expected to vanish outside the detector, hence $\alpha=0$. Naturally, this would spare us the technical problems posed by the initial conditions lying outside the detector. However, the time evolution of such compactly supported wave packets gets quite messy, which is avoided here for simplicity.}  Equation \eqref{split} can thus be written as
\begin{equation}\label{split2}
    \Lambda(\nu)=\frac{4}{\sqrt{\pi}}\int_0^d\!\!\mathrm{d}R_0~R_0^2\,e^{-R_0^2}\,\delta\!\left(\frac{1}{\sqrt{(d/R_0)^2-1}}-\nu\right)+\alpha(d)\,\delta(\nu).
\end{equation}
In order to evaluate the remaining integral in \eqref{split2}, we recall a useful identity of the Dirac delta function:
\begin{equation}\label{idd}
    \delta(f(x))=\sum_n\frac{\delta(x-x_n)}{|f'(x_n)|},
\end{equation}
where $x_n$ is a zero of the function $f$ (an $x$ for which $f(x)=0$), $f'$ denotes its derivative, and the sum runs over all (real) zeros of $f$. Choosing
\begin{equation}
    f(R_0)=\frac{1}{\sqrt{(d/R_0)^2-1}}-\nu,
\end{equation}
we obtain two zeros, viz.,
\begin{equation}
    R_0^{\pm}=\pm\frac{\nu}{\sqrt{1+\nu^2}}d,
\end{equation}
and evaluating the derivatives of $f$ at $R_0^{\pm}$, we find (applying \eqref{idd}):
\begin{equation}
    \delta\!\left(\frac{1}{\sqrt{(d/R_0)^2-1}}-\nu\right)=\frac{|R_0|^3}{d^2\nu^3}\Big(\,\delta(R_0-R_0^+)\,+\,\delta(R_0-R_0^-)\,\Big).
\end{equation}

\newpage

\begin{figure}[!ht]
\centering
\begin{overpic}[scale=1]{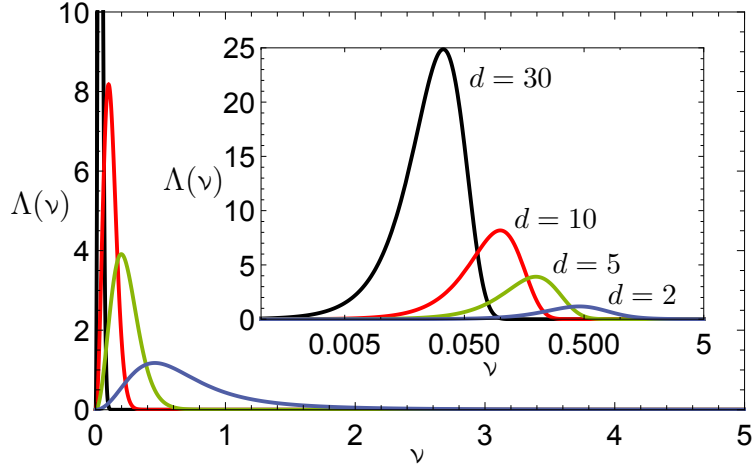}
\put (59,52) {$d=30$}
\put (65.5,31.5) {$d=10$}
\put (71,25) {$d=5$}
\put (79,20.5) {$d=2$}
\put (50.5,-2.5) {\large$\nu$}
\put (61,10) {\large$\nu$}
\put (15,37) {\large$\Lambda(\nu)$}
\put (-7,33.5) {\large$\Lambda(\nu)$}
\end{overpic}
\caption{Graphs of the continuous part of $\Lambda(\nu)$ vs. $\nu$ for different values of $d$. Inset: log-linear plot of the same curves.}
\label{fig1}
\end{figure}

\noindent Since $R_0^-<0$, only the first delta function term fires in the region of integration (cf Eq. \eqref{split2}), the integral is thus easily evaluated. After a few rounds of simplification, we obtain
\begin{equation}\label{gold}
    \Lambda(\nu)=\frac{4d^3}{\sqrt{\pi}}\,\frac{\nu^2}{(1+\nu^2)^{5/2}}\,\exp\!\left(-\frac{\nu^2}{1+\nu^2}\,d^2\right)+\alpha(d)\,\delta(\nu).
\end{equation}
We shall refer to the first (second) summand above as the \emph{continuous} (\emph{singular}) part of $\Lambda(\nu)$. The continuous part has been graphed in Fig. \ref{fig1} for different values of $d$.

\par
\begin{figure}[!ht]
\centering
\begin{overpic}[scale=1]{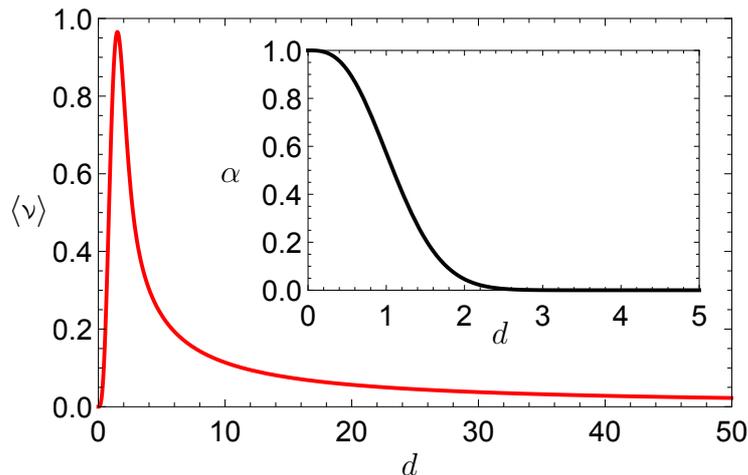}
\put (63,14.5) {\large$d$}
\put (24,38) {\large$\alpha$}
\put (50,-4.5) {\large$d$}
\put (-6.5,32.6) {\large$\expval{\nu}$}
\end{overpic}
\caption{Graph of mean reciprocal passage time $\expval{\nu}$ vs. detector radius $d$ (cf Eq. \eqref{mean}). Inset: Graph of $\alpha$ vs. detector radius $d$ (cf Eq. \eqref{alpha}).}
\label{fig2}
\end{figure}

 The reciprocal passage time distribution $\Lambda(\nu)$ has many interesting properties. First, we see from Fig. \ref{fig1} that the continuous part of the distribution becomes sharply peaked as $d$ becomes large, which implies that the first passage time distribution $\Pi(\tau)$ broadens with increasing $d$. In fact, as $d\to\infty$, $\alpha\to0$ (see Fig. \ref{fig2} inset), hence the singular part of $\Lambda(\nu)$ vanishes, while the continuous part tends to $\delta(\nu)$. This is reasonable, since the particle takes an infinite amount of time to cross a detector placed at infinity, hence the reciprocal passage time $\nu=0$ for all Bohmian trajectories.

\noindent On the other hand, as $d\to0$, i.e. as the detector shrinks to a point, all initial positions of the particle fall outside the detector volume, hence it never crosses the detector, consequently $\tau=\infty$ (or $\nu=0$) for all Bohmian trajectories and $\Lambda(\nu)\to\delta(\nu)$. The same conclusion follows from formula \eqref{gold}, since the continuous part now vanishes as $d\to0$, while $\alpha\to1$. Remarkably, in both limits $\Lambda(\nu)$ approaches a delta function, although for very different physical reasons.
\par Now, keeping $d$ fixed, we find that the continuous part of $\Lambda(\nu)$ grows as a power law $\sim\nu^2$ as $\nu$ approaches zero, and falls off as an inverse cube:
\begin{equation}
    \Lambda(\nu)\sim\frac{4d^3}{\sqrt{\pi}}\,\frac{e^{-d^2}}{\nu^3}+\mathcal{O}\big(\nu^{-4}\big)
\end{equation}
as $\nu\to\infty$. This implies that the mean reciprocal passage time $\expval{\nu}$ exists, i.e. it is finite, while its variance (or any higher cumulant) does not. Such distributions are said to be `heavy tailed'. We can also calculate $\expval{\nu}$ explicitly:
\begin{align}
    \expval{\nu}&=\int_0^{\infty}\!\!\!\mathrm{d}\nu~\nu\,\Lambda(\nu)=\frac{4d^3}{\sqrt{\pi}}\int_0^{\infty}\!\!\!\mathrm{d}\nu\,\frac{\nu^3}{(1+\nu^2)^{5/2}}\,\exp\!\left(-\frac{\nu^2}{1+\nu^2}\,d^2\right)\,+\,\alpha(d)\!\int_0^{\infty}\!\!\!\mathrm{d}\nu~\nu\,\delta(\nu)\nonumber\\
    &=\frac{2\,d^3}{\sqrt{\pi}}e^{-d^2}\int_0^1\!\!\!\mathrm{d}x~x^{-1/2}(1-x)\,e^{d^2x}\,+\,0\qquad\quad(\text{substituting $x=(1+\nu^2)^{-1}$})\nonumber\\
    &=\frac{2\,d^3}{\sqrt{\pi}}e^{-d^2}\times\frac{\Gamma(2)\Gamma\!\left(\frac{1}{2}\right)}{\Gamma\!\left(\frac{5}{2}\right)}\,_1F_1\!\left(\frac{1}{2};\frac{5}{2};d^2\right)\qquad\quad(\text{see Eq. (9.11.1) of \cite{Lebedev}})\nonumber\\&=\frac{8\,d^3}{3\sqrt{\pi}}\,_1F_1\!\left(\frac{1}{2};\frac{5}{2};d^2\right)e^{-d^2},\label{mean}
\end{align}
where $_1F_1(a;b;z)$ is the confluent hypergeometric function of the first kind.\footnote{In Ref. \cite{Lebedev} $_1F_1(a;b;z)$ is denoted by $\text{\textPhi}(a,b;z)$.} We graph Eq. \eqref{mean} in Fig. \ref{fig2}. Note that $\expval{\nu}$ vanishes as $d\to0$, and as $d\to\infty$, which is consistent with our earlier observations.
\section{Conclusion}
In the framework of Bohmian theory, we have derived the empirical first passage time distribution of a free particle, which has a satisfactory physical interpretation. Generally, these distributions depend strongly on the initial wave function of the particle. However, some of the features discussed here, for instance the behavior of $\Lambda(\nu)$ as $d\to\infty$ is rather universal. Therefore, in order to observe our results in real experiments one must prepare the initial wave function as accurately as possible.
\par A simple preparation procedure for realizing a desired $\psi_0$ was outlined by W. E. Lamb in \cite{Lamb}. The basic idea involves 1.) setting up a potential well $V(\bb{r})$ in some region of space, with $\psi_0$ being the ground state wave function of $V$, 2.) directing the particle from a source to this region, and 3.) waiting for radiation damping (or spontaneous emission) to bring the particle to the ground state. In the final step 4.), the potential $V$ is switched off \emph{suddenly}, allowing the particle to propagate freely in space. If the switching off is sufficiently fast, $\psi_0$ is left undisturbed. For preparing the initial wave function \eqref{init} we can choose $V(\bb{r})=\frac{1}{2}m\omega^2r^2$--a three dimensional isotropic harmonic potential--whose ground state wave function is a well known Gaussian (see \S\,13.2 of \cite{PictureBook}), which exactly equals \eqref{init} with $a=\sqrt{\frac{\hbar}{m\omega}}$. Therefore, appropriately tuning the trapping frequency $\omega$, one can fix the width $a$ to any desired value.
\par A final remark concerning the implications of our results is in order. Formula \eqref{singular} (or \eqref{lamb}) generally yields results different from other approaches, hence the possibility of experimentally distinguishing various proposals exists. With state of the art experimental technology, such as attosecond spectroscopy, our proposals might be checked in future experiments. ``\emph{Although empirical confirmation of these predictions would not prove the `reality' of the particle trajectory, it would provide strong circumstantial evidence in its favour, being a test of the particle law of motion}" \cite{Holland}.
\pagebreak
\section*{\small Acknowledgements}
\noindent\small I would like to thank Dr. J. M. Wilkes for critically reviewing this manuscript. The presentation of the paper improved greatly from Dr. Mike's suggestions. Dr. S. D. Mahanti inspired me to transcribe my seminar talk in the form of a paper for the Student Journal of Physics of IAPT. I thank him for his kind encouragement and continued support throughout the process. This paper is dedicated to Dr. Prof. Detlef D\"{u}rr for introducing me to this fascinating subject and for being a benevolent research partner ever since.

\end{document}